# A unified theory of grain growth in polycrystalline materials

**Authors:** Jianfeng Hu[1]*, Xianhao Wang[2], Junzhan Zhang[3,4], Zhijian Shen[4], Jun Luo[1,5], Jian Luo[6]

**Affiliations:**

[1] School of Materials Science and Engineering, Shanghai University, Shanghai 200444, China.

[2] Carl Zeiss Co., Shanghai, 60 Meiyue road, Shanghai 200131, China.

[3] Department of Materials and Environmental Chemistry, Arrhenius Laboratory, Stockholm University, S-106 91 Stockholm, Sweden.

[4] College of Materials and Mineral Resources, Xi'an University of Architecture and Technology, 13 Yanta Road, Xi'an 710055, PR China.

[5] Materials Genome Institute, Shanghai University, Shanghai 200444, China.

[6] Department of Nano Engineering, Program of Materials Science and Engineering, University of California, San Diego, La Jolla, CA 92093, USA

*Correspondence to:  jianfenghu0611@gmail.com

**Abstract**: Grain growth is a ubiquitous and fundamental phenomenon observed in the cellular structures with the grain assembly separated by a network of grain boundaries, including metals and ceramics. However, the underlying mechanism of grain growth has remained ambiguous for more than 60 years. The models for grain growth, based on the classically linear relationship between the grain boundary migration and capillary driving force, generally predict normal grain growth. Quantitative model for abnormal grain growth is lacking despite decades of efforts. Here, we present a unified model to reveal quantitatively how grain growth evolves, which predicts the normal, abnormal and stagnant behaviors of grain growth in polycrystalline materials. Our model indicates that the relationship between grain boundary migration and capillary driving force is generally nonlinear, but will switch to be the classically linear relationship in a specific case. Furthermore, the grain growth experiments observed in polycrystalline SrTiO$_3$ demonstrates the validity of the unified model. Our study provides a unified, quantitative model to understand and predict grain growth in polycrystalline materials, and thus offers helpful guides for the microstructural design to optimize the properties of polycrystalline materials.

**Main text**

The practical performances of polycrystalline materials are strongly affected by the formed microstructure inside, which is mostly dominated by grain growth behaviors [1–3]. For nanocrystalline materials, controlling grain growth is especially important to obtain and maintain the unique properties in nanoscale sizes[4–6]. Grain growth has been extensively investigated to understand grain growth behavior and control the microstructure for more than 60 years[7–16]. A linear relationship between the velocity of a grain-boundary migration and capillary driving force is generally used to model grain growth [17,18]. Based on this relationship, many quantitative equations (e.g. von Neumann-Mullins relation of topological model and a power growth law of mean-field model) were proposed to predict grain growth behaviors [8–10,17,19]. These models usually predict grain growth as a continuous process maintaining unimodal grain size distribution (uni-GSD), *i.e.* so-called normal grain growth (NGG) in polycrystalline materials [20–22]. However, besides NGG the microstructural evolution is frequently observed to be a discontinuous process with the occurrence of a bimodal grain size distribution (bi-GSD), *i.e.* abnormal grain growth (AGG), which still lack a well-accepted equation[23,24]. Many AGG mechanisms, including solute drag effect or particle/pore pinning at GBs [25–27], complexion transition [11–13], GBs faceting effect [14–16], have been proposed, each qualitatively valid in certain regimes. In contrast to the assumption of uniform GB features in NGG models, most of mechanisms attribute AGG to the existence of anisotropic GB features that is commonly believed to result in the preferential growth of a few grains with some special growth advantage over their neighbors in polycrystalline materials. In addition, although it is well accepted that the rate-limiting parameters like GB features and GSD play an important role on grain growth kinetics, a quantitative model incorporating them has yet to be established for both NGG and AGG in the past decades. Therefore, the underlying physical mechanism of grain growth has remained an unsolved problem for more than half a century in materials science [17,23].

**Model and Discussion**

Here, we consider grain growth at the atomic scale, since GB migration leading to grain growth results from atoms jumping across the interface. The rate at which a boundary moves is proportional to the net rate of atoms jumping across the boundary. According to a classical atomistic model of grain growth (see Fig. s1) [28,29], the rate of boundary migration can be given by

$$v = \lambda w_0 e^{\left(-\frac{\Delta G^*}{K_B T}\right)} [1 - e^{\left(-\frac{\Delta G}{K_B T}\right)}] \qquad 1,$$

where $\lambda$ and $w_0$ are the distance of each forward jump and the frequency of atomic jumps, respectively. $K_B$ and $T$ are the Boltzmann's constant and the temperature, respectively. $\Delta G^*$ is the apparent activation energy (or free-energy barrier) for atoms jumping across the boundary. $\Delta G$ is the difference in the free energy of material on the two sides of a grain boundary, which is the driving force for atomic transport in cellular microstructure. A derived process of classical model is provided in *SI Appendix* section 1. However, Eq.1 is very general and not specific enough to allow prediction of grain growth behaviors. Here, we develop further this model to achieve an exact equation for the quantitative prediction of grain growth. GB migration mostly involves two kinetic processes, *i.e.* the net detachment of atoms from the shrinking grain into GB region and net attachment of atoms from GB region onto the surface of growing grain, as shown in Fig. 1. Therefore, we suppose that the apparent activation energy $\Delta G^*$ involves the activation energy $\Delta G_{det}^*$ of atom detachment (or atom diffusion) and the free-energy barrier $\Delta G_{att}^*$ for the atom attachment to the growing grain. The atom-detached process is a thermally activated process, *i.e.* $\Delta G_{det}^* = \Delta Q$. The detached atoms in boundary region have the kinetic energy of $K_B T$ according to equipartition theorem, and thus obtain the frequency of atomic jumps $w_0 = \frac{K_B T}{h}$ according to de Broglie relations, where $h$ is the Plank's constant. On the other hand, the energy barrier of the nucleation limits mostly the process of atom attachment onto the surface of growing grain. According to crystal growth theory[30], the energy barrier of the nucleation on the crystal surface is usually related to the step free energy, which is determined not only by the crystal face and the crystal structure but also to the surface roughness. The surface roughness increases continuously with approaching the surface-roughening transition (from smooth surface to rough surface on the atomic scale), which is accompanied by the decrease in step free energy until zero, *i.e.* no energy barrier to continuous growth, at the roughening transition. Similarly, we suppose that the energy barrier of atomic attachment onto growing grain is also related to a step free energy at GB that equivalent to step free energy at crystal surface. Therefore, the free energy change $\Delta G_{att}$ to form a new layer of height $l$ and radius $R$ on growing-grain surface is equal to the energy difference between the increase in the free energy forming a new layer ($2\pi R \varepsilon$) and the decrease in GB free energy ($\pi R^2 l \cdot \Delta P$), *i.e.* $\Delta G_{att} = 2\pi R \varepsilon - \pi R^2 l \cdot \Delta P$. Here, $\varepsilon$ is defined as the free energy per unit length of the perimeter of new layer on the surface of growing grains, analogous to step free energy in crystal growth theory. The more recent atomic-scale observations of GB features revealed that the atomic structure of solute segregation was dominated by the crystallography of the terminating surface of grains rather than GB mis-orientation as commonly believed[31,32], which is consistent with the assumption of atom-attached nucleation process for grain growth in our model. Compared with step free energy in crystal growth theory, $\varepsilon$ is related not only with crystallographic features of growing grains and temperature but also with the surrounding GB features (e.g. defects and compositions). $\Delta P$ is the capillary pressure difference caused by the surface curvatures on the two sides of a grain boundary, *i.e.* $2\gamma(\kappa_a - \kappa)$. Here, $\gamma$ is the boundary energy. $\kappa$ and $\kappa_a$ are the mean curvatures of growing grain and shrinking grain on the two side of the GB , respectively, which are statistically related with the corresponding grain sizes in polycrystalline system. Unlike the assumption of $\kappa = -\kappa_a$ in classical models (see Fig.s1)[28,29], we suppose that $\kappa$ and $\kappa_a$ can be independent for each other and there has $\kappa < \kappa_a$ in the case of grain growth. Therefore, we can further obtain the critical energy barrier of a new layer as $\Delta G_{att}^* = \frac{\pi \varepsilon^2}{2l\gamma(\kappa_a-\kappa)}$, which is given by the maximum value of $\Delta G_{att}$. Since the driving force results from the surface curvature difference between adjacent grains on both sides of the GB, we can write their difference in the free energy as $\Delta G = 2\gamma\Omega(\kappa_a - \kappa)$ (the Gibbs-Thompson effect). $\Omega$ is the atomic volume. Due to the value of $\Delta G$ usually far smaller than that of $K_B T$, i.e. $\Delta G \ll K_B T$, the term of $1 - e^{(-\frac{\Delta G}{K_B T})}$ in Eq. 1 is approximately equal to $\frac{\Delta G}{K_B T}$, i.e. $1 - e^{(-\frac{\Delta G}{K_B T})} \cong \frac{\Delta G}{K_B T}$. Now, we can rewrite Eq. 1 as

$$v = M\gamma\kappa(n-1)e^{\left(-C \cdot \frac{\varepsilon^*}{T(n-1)\kappa}\right)} \qquad 2$$

where, $M = \frac{2\lambda\Omega}{h} e^{\left(-\frac{\Delta Q}{K_B T}\right)}$ is the temperature-dependent coefficient that can be regarded as GB mobility, analogous to the counterpart in the classical models. $n = \frac{\kappa_a}{\kappa}$ is a grain-size-related dimensionless variable, since the mean curvature of polyhedral grains at GB is closely associated with grain sizes. Therefore, the value $n$ is statistically determined by grain sizes and its distribution in polycrystalline materials. $C = \pi/2K_B$ is the constant. $\varepsilon^*$ is defined as $\varepsilon^* = \varepsilon^2/l\gamma$ and named the GB step free energy due to the same dimension with the step free energy in crystal-growth theory, i.e. the free energy per unit length. According to the definition, the GB step free energy is jointly determined by the crystallographic features of terminating surface of growing grain and its surrounding GB conditions (e.g., segregations and defects at GB). Analogous to crystal growth, GB roughening, caused by increasing temperature or changing GB chemical compositions [33], can result in structure transition from singular surface to rough surface that accompanied by the decrease in $\varepsilon^*$ until zero. Therefore, the variable $\varepsilon^*$ reflects the difference in energy barriers of the atom attachment among distinct GB nucleation environments.

The general rate of change of the volume $V$ of individual grain (polyhedron with N faces) is equal to the summation of the increase in volumes over all faces of

polyhedron, which is achieved by GB migrations and can thus be given by:

$$\frac{dV}{dt} = \sum_{i=1}^{N} v_i \cdot A_i = \sum_{i=1}^{N} A_i \cdot M\gamma_i \kappa_i (n_i - 1) e^{\left(-C \cdot \frac{\varepsilon_i^*}{T(n_i-1)\kappa_i}\right)} \qquad 3.$$

Here, $v_i$ and $A_i$ are the migration rate and the area of the $i$th face of growing grain, respectively. Therefore, Eq. 3 can depict the topological evolution of individual grains during growth in polycrystalline materials. According to Eq. 3, grains with anisotropic GB energy and GB step free energy may result in anisotropic grain growth, such as the rod-like grains in polycrystalline $Si_3N_4$ materials and the plate-like grains in polycrystalline $Al_2O_3$ materials. In addition, as mentioned above, GB roughening transition caused by temperature or/and chemical segregations will result in the case of $\varepsilon_i^* = 0$. For this specific case of $\varepsilon_i^* = 0$ and the uniform GB energy over all boundaries in polycrystalline systems, Eq. 3 will reduce to

$$\frac{dV}{dt} = \sum_{i=1}^{N} A_i \cdot M\gamma\kappa_i (n_i - 1) \qquad 4.$$

In this case, there is a linear relationship between grain growth rate $\frac{dV}{dt}$ and the capillary driving force $\gamma\kappa(n-1)$, which is same with the classically theoretical treatments for NGG[8–10,17,19–21]. Compared with von Neumann-Mullins relation, Eq.4 incorporates the variable $n$ to reveal the effect of size difference between adjacent grains on the growth of individual grains that accords with Ostwald ripening. Furthermore, compared to Eq.3, Eq.4 reveals that NGG is a specific case of grain growth in polycrystalline materials.

A major limitation in studying grain growth is that it is hard to probe the real-time, three-dimensional evolution of grains during heating treatment. In practice, the size of a polyhedral grain in polycrystalline materials is generally described by an average caliper dimension (linear size) called mean caliper diameter, *i.e.* the mean perpendicular distance (averaged over all orientations) between two parallel tangent planes on the polyhedron[34]. In order to assess the validity of Eq. 3 for grain growth, we need to simplify the rate of change of (three-dimensional) volume into the growth rate of (one-dimensional) linear size of grains for the experimental verifications. According to Minkowski's theorem, the average mean curvature of grain is inversely related with its mean caliper diameter [35,36]. For the case of isotropic grains with uniform GB step free energy $\varepsilon^*$ and GB energy $\gamma$, therefore, the growth rate in Eq. 3 can be approximately

simplified by a linear characteristic as

$$\frac{dD}{dt} = M\gamma(n-1)\frac{1}{D} \cdot exp\left(-C \cdot \frac{\varepsilon^* D}{T(n-1)}\right) \qquad 5,$$

where, $D$ is the mean caliper diameter (also linear size) of polyhedral grains. A derived process is provided in *SI Appendix* section 1. The dimensionless variable $n$ is then equal to the ratio of $D$ of growing grain to the average mean caliper diameter $D_a$ of adjacent smaller grains, *i.e.* $n = D/D_a$. Therefore, the dimensionless variable $n$ is usually in the value range of 1 to 4 in polycrystalline materials with uni-GSD. Analogous to Eq.4, Eq.5 will reduce to $\frac{dD}{dt} = M\gamma(n-1)\frac{1}{D}$ in the specific case of $\varepsilon^* = 0$. The specific equation is similar with the Hillert's classical theory[9] for NGG, except for the growth model of individual grains instead of the mean-field model of Hillert's theory. On the base of Eq.5, we used numerical simulations to illustrate quantitatively the effects of important variables on grain growth in order to reveal the evolution of grain growth behavior in polycrystalline materials. The effects of $\varepsilon^*/T$ with $n$ and $D$ with $n$ on the growth rate of individual grains are quantitatively illustrated in Fig. 2, respectively. More numerical simulations in details see in *SI Appendix* section 2. Figure 2 reveals that the existence of growth switch of individual grains between growth and stagnation, which switch is together dominated by the variables of $\varepsilon^*/T$, $n$ and $D$. This numerical simulation also indicates that grain-growth switch is mostly dominated by the exponential term in Eq. 5. The temperature-dependent variable $\varepsilon^*$ decreases with increasing temperature as mentioned above, $\varepsilon^*/T$ decreases even faster than $\varepsilon^*$ with increasing temperature. The decrease of $\varepsilon^*/T$ or the increase of $n$ may result in the exponential increase in growth rate of individual grains, and thus may lead to the switch from stagnation to rapid growth as shown in Fig.2a. Analogous to the decrease in $\varepsilon^*/T$ with increasing temperature, the increase in $D$ is also energetically favorable trends during increasing temperature or holding time at elevated temperature. Figure 2b illustrates the effects of $D$ with $n$ on grain growth rate, which also indicates the existence of the growth switch of individual grains. In contrast to the effect of $\varepsilon^*/T$ with increasing temperature on grain growth, the increase of $D$ value may result in the growth switch of individual grains from rapid growth to stagnation as shown in Fig.2b.

In contrast to our common belief of larger grains preferred to grow in a grain size distribution, larger grains with low $n$ may have lower growth rate than that of smaller grains with high $n$ in polycrystalline system. The numerical simulations in Fig. 2 reveal that a growth rate distribution (GRD) corresponding to a GSD may exist in polycrystalline systems. Three kinds of GRDs may occur in polycrystalline systems, *i.e.* GRD with only rapid-growth grains, GRD with only stagnant-growth grains, or GRD with the coexistence of rapid-growth grains and stagnant-growth grains. The case of only rapid-growth grains existing in polycrystalline materials may occur in the case of zero GB step free energy ($\varepsilon^* = 0$), which leads to NGG according to Eq. 4. The existence of only stagnant-growth grains may result in grain growth stagnation (GGS) in polycrystalline materials. In the third case, the coexistence of rapid-growth grains and stagnant-growth grains may result in the occurrence of a few abnormal grains (rapid-growth grains) in smaller grain matrix (stagnant-growth grains) in polycrystalline materials, *i.e.* AGG. Numerical computer simulations also indicated that AGG occurred if part of grains had some sorts of growth advantages over the other part of grains in polycrystalline materials[37,38]. In contrast to the assumed growth advantages like preferred surface energy or GB mobility in computer simulations, this work illustrates that AGG may intrinsically occur due to the difference in grain sizes as shown in Fig. 2b. It also reveals that grain growth behavior may become similar with that of NGG if the value of $\varepsilon^*/T$ approaches to zero as shown in Fig.2a, which may be the reason of the existence of various growth exponents for NGG as commonly observed in numerous grain growth experiments[15,17]. Therefore, the derived equations can successfully predict how grain growth evolves, which including AGG and GGS besides NGG in polycrystalline materials.

**Experiment and Simulation**

We further designed a grain growth experiment to verify this growth theory. In order to simplify the effects of experimental factors on grain growth, we chose a model SrTiO$_3$ system without additives and with relatively uniform cubic-faceted nanocrystals as the starting powder. Meanwhile, the high cooling rate of more than 320 °C/min was obtained by using sparking plasma sintering (SPS) technology to freeze the high-temperature microstructures of samples as much as possible. The experimental results reveal the alternate occurrence of AGG and GGS during increasing heating temperatures as shown in Fig.3, which similar phenomenon was also observed in a nickel system[39]. The GSD of sample at 900 °C with a holding time of 3 minutes shows clearly two peaks, *i.e.* bimodal GSD with the partial overlap between two size distributions, which means the occurrence of AGG at 900 °C. Therefore, AGG occurred at 900 °C and 1000 °C, while GGS occurred at 950 °C and 1100 °C. Furthermore, GSD of matrix grains in AGG samples remained relativelt stationary during the rapid growth of abnormal grains as shown in Fig. 3a. The growth of abnormal grains maintained the same cubic-faceted morphology with the starting $SrTiO_3$ nanocrystals up to micro-scale size as shown in Fig.3c-3d. This indicates that the growth of cubic-faceted $SrTiO_3$ crystallites were accomplished by the migration of uniform {100} planes through GB during heating treatment at elevated temperatures.

The growth behaviors in the $SrTiO_3$ samples can be quantitatively analyzed by the derived equations. The cubic-faceted $SrTiO_3$ grains have a uniform GB step free energy $\varepsilon^*$ at each specific temperature due to the uniform {100} planes as surfaces during grain growth. The evolution of grain growth behaviors can be interpreted by using Eq.5. First, the initial grain-growth rate distributions of samples at the onset of different temperatures are simulated to illustrate in Fig. 4. The calculated results reveal that the existence of growth switch divides grains into rapid-growth grains (*i.e.* non-zero growth rate) and stagnant-growth grains (*i.e.* zero growth rate) in the samples at the onset of 900 °C (Fig. 4a) and 1000 °C (Fig. 4c), respectively. The switch of grain growth is mostly related with the combination of $n$ value and grain sizes. Second, according to Eq.5, the rapid-growth grains would increase their sizes at the expense of the adjacent smaller grains (stagnant-growth grains), and then accelerate their growth rate in the initial stages due to the increase in $n$ as illustrated in Fig.2b and Fig. s2. Meanwhile, the stagnant-growth grains as the matrix grains maintained grain-growth stagnation or were consumed by the growing grains with increasing time. Hence, AGG occurred in these two samples with the coexistence of rapid-growth grains and stagnant-growth grains. With prolonging holding time and consuming rapidly the matrix grains, the growth rates of abnormal grains would in turn decrease due to the increase in grain size and decrease in the $n$ values among grown grains as predicted in Fig. 2. The reduction of $n$ value results from the impingement of growing grains in Fig. 3c due to decreasing the amount of matrix grains. Last, abnormal grains would stop growing and thus GGS occurred after the matrix grains were consumed up. Meanwhile, bimodal GSD would vary to unimodal GSD as observed in samples at 950 °C and 1100 °C in Fig. 3. Furthermore, when increasing temperature of simples, the $\varepsilon^*$ value will decrease and thus induce the growth transition of individual grains from stagnation to growth as illustrated in Fig. 2a, which resulted in the transition of grain growth behavior from GGS to AGG as observed in the samples at 950 °C and 1000 °C. Therefore, the Eq. 5 well interprets the alternate occurrence of AGG and GGS observed in this polycrystalline SrTiO$_3$ materials.

Controlling grain size is fundamentally important for obtaining and maintaining the high performances of polycrystalline materials, especially for the nanocrystalline materials. Although GGS plays a crucial role on controlling grain size, the underlying mechanism remains ambiguous. The occurrence of GGS is usually ascribed to the pinning of solutes[25] or second-phase particles at GBs[40,41]. More recently, the smooth-boundary pinning was proposed to result in GGS in pure polycrystalline materials[42]. Each of these phenomenological interpretations is only valid in certain cases. In this work, we quantitatively demonstrate that the occurrence of GGS is synergistically dominated by the intrinsic variables of GB step free energy, grain size and its distribution as shown in Fig. 2, which may be a general mechanism for polycrystalline materials.

**Summary**

A unified theory is presented to successfully predict and depict grain growth behaviors including NGG, AGG and GGS in polycrystalline materials, which involves the variables of GB step free energy, grain size and its distribution. The variable of GB step free energy $\varepsilon^*$ is introduced to reflect the effects of the morphology of growing grain and surrounding GB feathers on grain growth, which value can be influenced by the experimental conditions like the additives and temperature. According to Eq. 3, a single $\varepsilon^*$ value can develop the isotropic morphology as the observation of cubic-shape $SrTiO_3$ grains, while grains with anisotropic $\varepsilon^*$ values may lead to the anisotropic morphology of grains. The derived formula reveals that ideal NGG only occurs in the specific case that $\varepsilon^*$ equal to zero. In a word, the unified theory offers a conclusive model to predict how grain growth evolves and interpret the grain growth behaviors in polycrystalline materials. Furthermore, it may allow us to accurately tailor and design the microstructures and properties of polycrystalline materials.

**Acknowledgments:**

We thank Prof. Hui Gu for fruitful discussions. J. H. designed the research; J.H., X.W., J.Z., Z.S., J.L. and J.L. performed experiments; J. H. analyzed the data and wrote the paper. All authors discussed the results and commented on the manuscript. The authors declare no conflicts of interest. The data reported in this paper are available from the authors.

**Figure Captions**

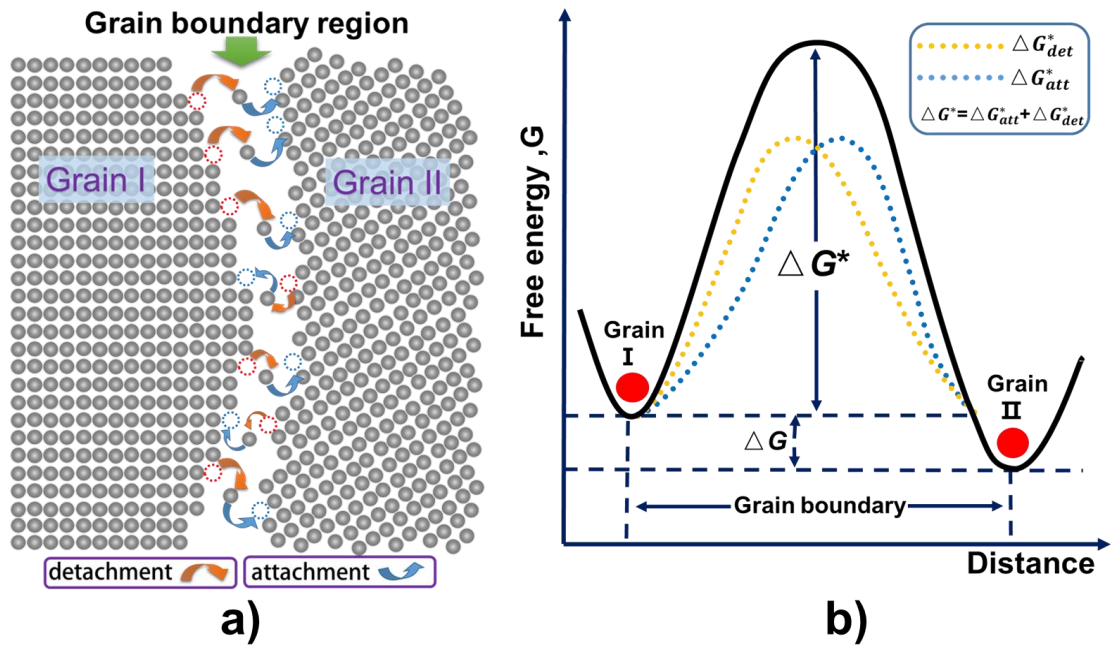

**Fig. 1. The schematic model of grain boundary migration by means of atom jumps.** **(a)** The mechanism of migration including two processes of atomic detachment and atomic attachment**; (b)** the free energy of an atom during a jump across the boundary.

The apparent activation energy $\Delta G^*$ of grain boundary migration involves the activation energy of atom detachment and energy barrier of atom attachment.

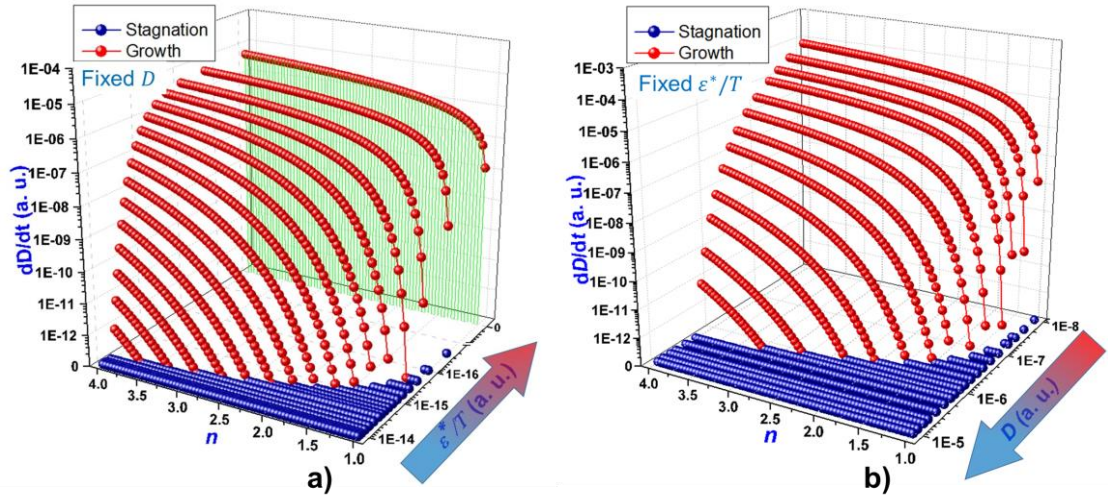

**Fig. 2. Schematic illustration of the growth rate of individual grain evolving with the changes of $\varepsilon^*/T$, $D$ and $n$ in Eq. 5. Grain growth rate divides grains into two states: growth and stagnation.** **(a)** 3D plot of grain growth rate as a function of $\varepsilon^*/T$ and $n$ for a fixed $D$ value. The value $\varepsilon^*/T$ can decrease with increasing temperature and goes to zero at GB transition temperature as indicated by arrow. The variable of $n$, i.e. $D/D_a$, is associated with grain size distribution in polycrystalline system. **(b)** 3D plot of grain growth rate as a function of $D$ and $n$ for a fixed $\varepsilon^*/T$ value. Grain size $D$ is energetically favorable to increase with increasing heating-treatment temperature or holding time (indicated by arrow). **(a)** and **(b)** illustrate the opposite effect of $\varepsilon^*/T$ and $D$ on grain growth rate when increasing temperature.

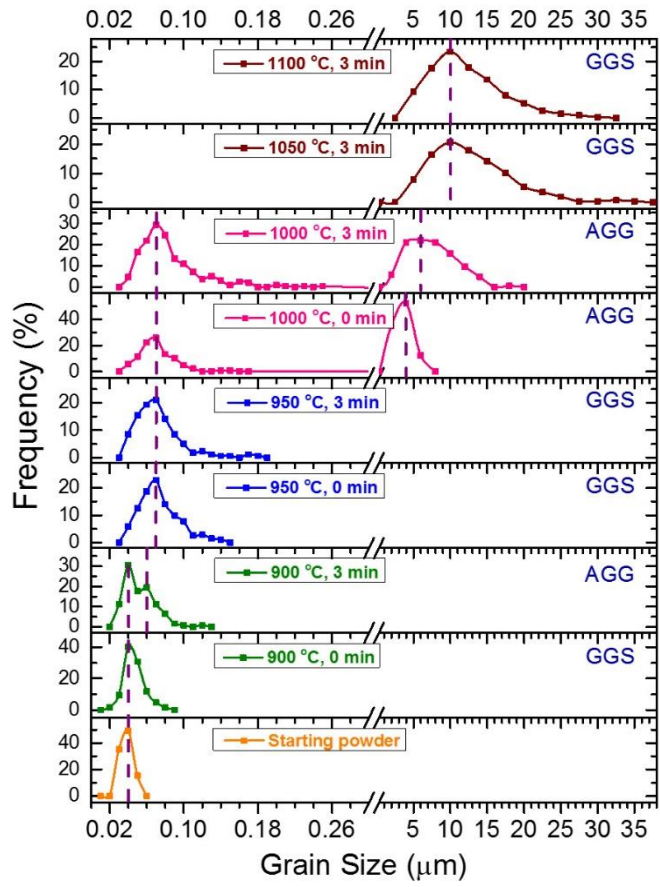

(a)

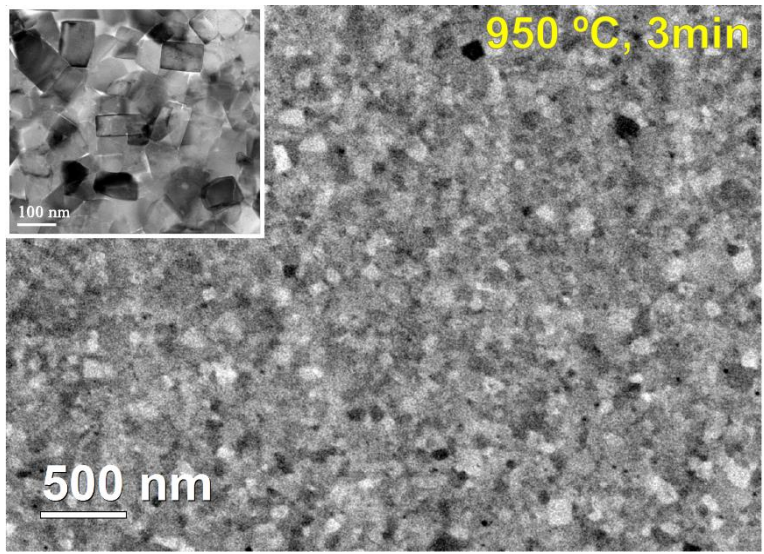

(b)

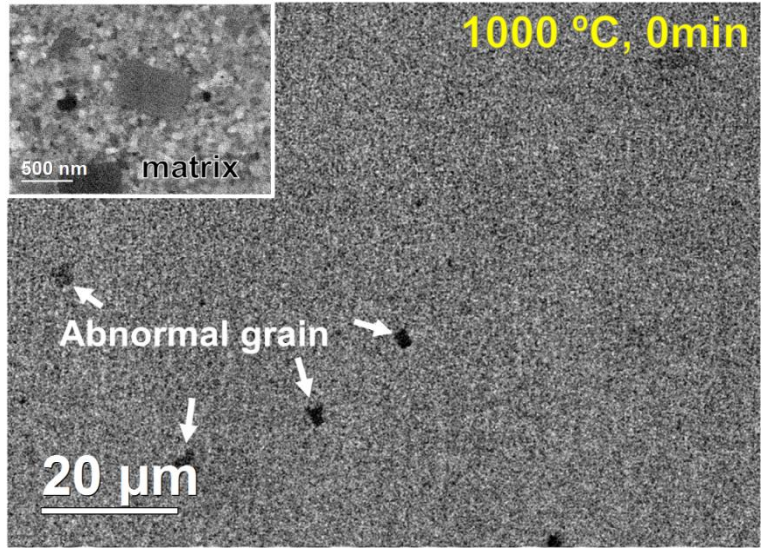

(c)

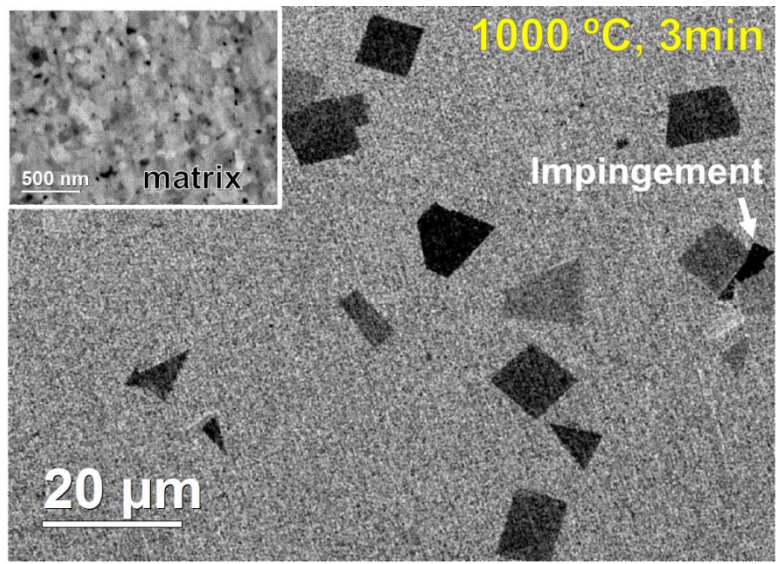

(d)

**Fig. 3**. **The evolutions of grain size distribution (GSD) and grain morphology in the sintered samples.** (**a**) The development of grain size distributions with increasing sintering temperature and holding time. The plot shows the alternate occurrence of abnormal grain growth (AGG) and grain growth stagnation (GGS) in sintered samples. The position of matrix GSD remains stationary during AGG as pointed out by dash lines. (**b**)-(**d**) Electron micrographs showing the grain morphologies in the sintered samples at different temperatures. (**c**) and (**d**) SEM micrographs showing submicron-

and micron-sized abnormal grains in sample sintered at 1000 °C with zero holding time and 3 minutes.

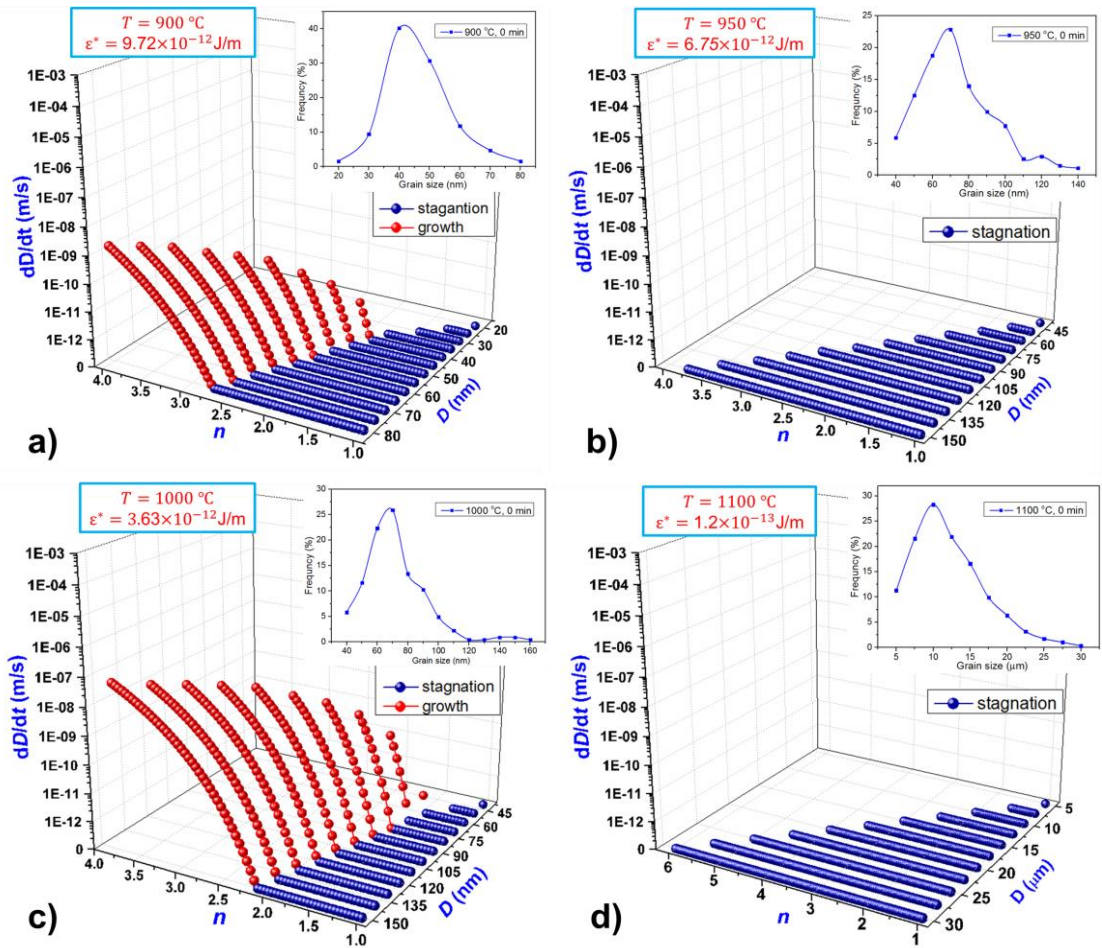

**Fig. 4**. **Probability distribution of grain growth rates at the onset of heating-treatment temperature in polycrystalline SrTiO₃.** The results of **a)** to **d)** are simulated by using Eq. 5 and experimental GSD data in Fig. 3 (insets on the upper right of figures). **a)** and **c)** growth rate distribution at the onset of 900 °C and 1000 °C. The simulated results reveal that these two samples involves rapid-growth grains and stagnant-growth grains inside, which will result in abnormal grain growth (AGG). (**b**) and (**d**) growth rate distribution at the onset of 950 °C and 1100 °C. All grains in these two samples remain stagnant growth.